# RF-Transformer: A Unified Backscatter Radio Hardware Abstraction

Xiuzhen Guo[1], Yuan He[1], Zihao Yu[1], Jiacheng Zhang[1], Yunhao Liu[2], Longfei Shangguan[3]
[1]School of Software and BNRist, Tsinghua University
[2]School of Global Innovation Exchange (GIX) and Department of Automation, Tsinghua University
[3]Department of Computer Science, University of Pittsburgh
guoxiuzhen94@gmail.com,heyuan@mail.tsinghua.edu.cn,{zh-yu17,zhangjc21}@mails.tsinghua.edu.cn,
yunhaoliu@gmail.com,longfei@pitt.edu

## ABSTRACT

This paper presents RF-Transformer, a unified backscatter radio hardware abstraction that allows a low-power IoT device to directly communicate with heterogeneous wireless receivers at the minimum power consumption. Unlike existing backscatter systems that are tailored to a specific wireless communication protocol, RF-Transformer provides a programmable interface to the micro-controller, allowing IoT devices to synthesize different types of protocol-compliant backscatter signals sharing radically different PHY-layer designs. To show the efficacy of our design, we implement a PCB prototype of RF-Transformer on 2.4 GHz ISM band and showcase its capability on generating standard ZigBee, Bluetooth, LoRa, and Wi-Fi 802.11b/g/n/ac packets. Our extensive field studies show that RF-Transformer achieves 23.8 Mbps, 247.1 Kbps, 986.5 Kbps, and 27.3 Kbps throughput when generating standard Wi-Fi, ZigBee, Bluetooth, and LoRa signals while consuming 7.6–74.2$\times$ less power than their active counterparts. Our ASIC simulation based on the 65-nm CMOS process shows that the power gain of RF-Transformer can further grow to 92–678$\times$. We further integrate RF-Transformer with pressure sensors and present a case study on detecting foot traffic density in hallways. Our 7-day case studies demonstrate RF-Transformer can reliably transmit sensor data to a commodity gateway by synthesizing LoRa packets on top of Wi-Fi signals. Our experimental results also verify the compatibility of RF-Transformer with commodity receivers. Code and hardware schematics can be found at: https://github.com/LeFsCC/RF-Transformer.

## CCS CONCEPTS

• **Networks** → **Network architectures**; • **Computer systems organization** → **Embedded and cyber-physical systems**.

## KEYWORDS

Wireless communication; Internet of Things (IoT); Backscatter technology; PHY layer design

---

Yuan He and Longfei Shangguan are the co-corresponding authors.

**ACM Reference Format:**
Xiuzhen Guo[1], Yuan He[1], Zihao Yu[1], Jiacheng Zhang[1], Yunhao Liu[2], Longfei Shangguan[3]. 2022. RF-Transformer: A Unified Backscatter Radio Hardware Abstraction. In *The 28th Annual International Conference On Mobile Computing And Networking (ACM MobiCom '22), October 17–21, 2022, Sydney, NSW, Australia.* ACM, New York, NY, USA, 13 pages. https://doi.org/10.1145/3495243.3560549

## 1 INTRODUCTION

The last decade has witnessed remarkable advance in backscatter technology [30, 31, 38, 42, 46, 49, 58, 59]. To save power, the backscatter radio (*a.k.a.*, passive radio) does not generate carrier signal; it instead modulates data on top of an incident carrier signal emanating from another wireless device called carrier signal generator. As the power consumption of communication is dominated by the carrier signal generation [34, 53], the backscatter radio thus consumes orders of magnitude lower power than the active radio, making it an appealing solution to low-power IoT devices.

The proliferation of Internet of Things (IoT) applications brings about the increasingly dense deployments of various wireless devices (*e.g.* Wi-Fi, ZigBee, Bluetooth, LoRa, etc.). The coexistence of heterogeneous wireless devices puts forward more stringent requirements for the adaptability and flexibility of the backscatter design. In order to integrate seamlessly into heterogeneous wireless networks, the backscatter radio should be able to interplay directly with different technologies while maintaining ultra-low power consumption.

However, most of existing backscatter systems are *tailored to* a specific wireless technology but lack *flexibility*, *i.e.*, synthesizing Wi-Fi [38, 39, 58] or LoRa [31, 46, 49][1], but not both because Wi-Fi and LoRa share radically different physical layer designs. Such a tailored backscatter design is resulting in low adaptability, obstructing its practical deployment in heterogeneous wireless networks.

To retain the flexibility of signal generation, the backscatter radio should be programmable, allowing IoT devices to synthesize different backscatter signals to adapt. However, such programmability is usually achieved by radio function softwarization [33] (*e.g.*, software-defined radios) which comes with significant computation overhead and energy footprint, contradicting the low-power principle of backscatter system.

In this paper, we ask the following question: *Is it possible to build a programmable backscatter radio that can generate different types of protocol-compliant wireless signals while retaining ultra-low power consumption?* A positive answer would pave the way to the practical deployment of backscatter systems in heterogeneous wireless networks. For instance, the backscatter radio synthesizes standard Wi-Fi packets when talking to a Wi-Fi Access Point (AP) but swiftly switches to generating standard Bluetooth or ZigBee signals for direct device-to-device communication when Wi-Fi APs are not available; The backscatter radio presents different capabilities in communication range and throughput to adapt to various scenarios, *i.e.* directly transmitting commands to the indoor Wi-Fi AP or the outdoor LoRa gateway; Moreover, since an IoT device equipped with the backscatter radio could directly communicate with a Wi-Fi/Bluetooth/ZigBee/LoRa receiver, we envision this backscatter radio could eliminate the use of gateways and thus alleviate wireless traffic congestion on the overcrowded unlicensed band.

---

[1] The latest LoRa chip SX1280 [2] can operate on 2.4 GHz freq. band.

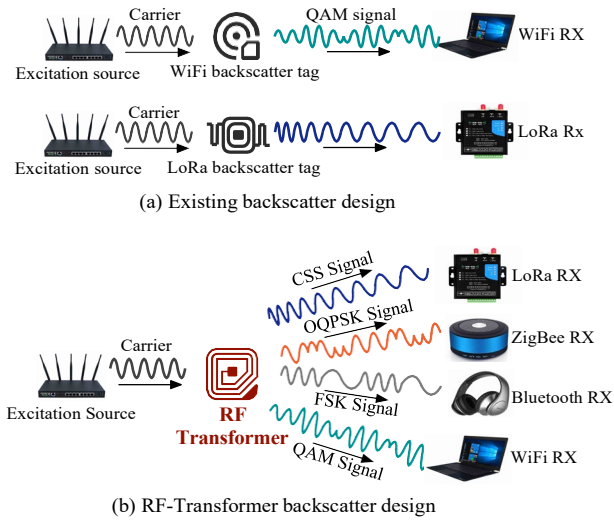

**Figure 1: Comparison of (a) existing backscatter systems and (b) RF-Transformer.** RF-Transformer retains both flexibility and small power footprint. It can synthesize different types of backscatter signals based on a unified radio hardware.

We give an affirmative answer by presenting RF-Transformer, a unified backscatter radio hardware abstraction that supports synthesizing different protocol-compliant wireless signals on top of a carrier signal emanating from either a helper device [38] or a commodity wireless device[2]. Our design is based on an observation that the change of reflection coefficient of a backscatter radio will alter both the phase and amplitude of backscatter (reflected) signals. This allows RF-Transformer to generate different amplitude-modulated and phase-modulated backscatter signals. Moreover, given that the frequency of a backscatter signal can be manipulated by changing its phase variation rate; thus it is also feasible to generate frequency-modulated backscatter signals with RF-Transformer.

The reflection coefficient of a backscatter radio is determined by its load impedance connected to the radio antenna. The load impedance comprises of *resistive* load and *reactive* load two parts, with the former determining the amplitude of backscatter signals while the latter controlling their phase. To generate any type of backscatter signal, it is crucial to respectively program the resistive and reactive load in fine granularity. However, achieving this goal while retaining the low-power nature of backscatter radio is challenging because the low-power and high programmability dual requirement requires us to offload the sophisticated control logic of load impedance to the analog domain, which will complicate the radio design; add its cost, weight, and form factor (§3.2).

To tradeoff, the current practice on backscatter design sacrifices the programmability to retain a small power footprint. For instance, Interscatter [35] allows the backscatter radio to generate different DBPSK or DQPSK symbols by toggling the RF switch among four different, pre-defined load impedances. Albeit low-power, Interscatter is limited to phase shift-based modulation and cannot scale to generate many other types of backscatter symbols due to its limited space on load impedance selection.

To retain both high programmability and low power footprint, RF-Transformer uses the insight that both phase and amplitude of backscatter signals can be programmed by altering the amplitude of their In-phase part and Quadrature part, respectively (§3.1). Hence the problem is transformed into *how to modulate the amplitude of the In-phase part and Quadrature part of backscatter signals*? To answer this question, we leverage a passive RF splitter coupled with a delay line to build In-phase path (I-path) and Quadrature path (Q-path) on backscatter radio and terminate each path with an ultra-low power MOSFET transistor. The resistive load of a MOSFET transistor changes with its bias voltage. As the amplitude of signal propagating along each path changes with its resistive load, we can program the bias voltage of these two MOSFET transistors to alter the amplitude of In-phase and Quadrature part of backscatter signals, thereby generating different types of backscatter signal.

**Implementation**. We prototype RF-Transformer on printed circuit board (PCB) hardware and experiment with software-defined radios in various scenarios. Our evaluation shows that RF-Transformer can generate 23.8 Mbps Wi-Fi signals, 247.1 Kbps ZigBee signals, 986.5 Kbps Bluetooth signals, and 27.3 Kbps LoRa signals at 7.6–74.2× lower power consumption compared against their active radios. We simulate the Application Specific Integrated Circuit (ASIC) of RF-Transformer based on TSMC 65-nm CMOS process and show that the power consumption can be further reduced to 371.2 $\mu$W, 80.1 $\mu$W, 89.2 $\mu$W, and 47.8 $\mu$W for Wi-Fi, Bluetooth, ZigBee, and LoRa generation, respectively.

We further demonstrate RF-Transformer can even modulate Wi-Fi traffic into standard LoRa packets. To show the potential of our design, we implement a proof-of-concept IoT device based on RF-Transformer for foot traffic density monitoring indoors. Our 7-day evaluation shows RF-Transformer can successfully transmit LoRa packets synthesized on top of Wi-Fi signals to a commercial LoRa gateway with the average packet reception ratio (PRR) of 90%. Our experimental results also demonstrate that the backscatter signals generated by RF-Transformer can be received by the commodity wireless transceivers on laptops and smartphones.

This paper makes the following contributions. *i*) We propose a low-power, unified backscatter radio hardware abstraction that can synthesize different types of backscatter signals. With this design, the IoT devices equipped with RF-Transformer can talk to heterogeneous wireless devices without relying on a dedicated gateway. To the best of knowledge, RF-Transformer is the first-of-its-kind backscatter radio design that achieves both low power and high flexibility. *ii*) We address both design and implementation challenges of RF-Transformer and demonstrate its efficacy through comprehensive experiments and long-term case studies. *iii*) We further demonstrate RF-Transformer can support cross-technology backscatter, *i.e.*, synthesizing LoRa packets on top of Wi-Fi transmissions. RF-Transformer is a significant step in a line of works that will scale out backscatter technologies to heterogeneous wireless networks.

---
[2]One can generate a carrier signal using commodity devices by manipulating the payload of their transmitted Bluetooth [35] or Wi-Fi [41] packet.

**Roadmap**. The rest of this paper is organized as follows. We review related works in Section 2. We present the design of RF-Transformer in Section 3, followed by practical considerations in Section 4. Section 5 describes cross-technology backscatter design. The implementation details (§6) and experiment (§7) follow. We conclude in Section 8.

## 2 RELATED WORKS

From the perspective of the modulation approach, existing backscatter systems can be divided into three groups.

**ON-OFF Keying**. The first group of works adopts ON-OFF Keying (OOK) [32] to modulate tag data, which is achieved by switching between *reflecting* and *absorbing* two states. OOK modulation has been widely used in passive RFID systems [34, 53]. Later on, Ambient Backscatter [42] and Turbocharging [45] take advantage of OOK modulation to backscatter ambient TV and cellular signals, respectively. Wi-Fi Backscatter [38] further extends OOK modulation to Wi-Fi traffic, allowing Wi-Fi receivers to demodulate backscatter signals by detecting the received signal strength variation. mmTag [44] employs OOK modulation for mmWave backscatter. Albeit simple and power efficient, OOK achieves a very low link throughput.

**Phase shifting**. The second group of works proposes phase modulation to improve link throughput and synthesize protocol-compliant backscatter symbols. BackFi [23] builds a set of delay lines with different lengths to enable BPSK, QPSK, and 16-PSK modulation. HitchHike [56] proposes codeword translation that allows a backscatter tag to encode tag data by translating the original Wi-Fi 802.11b codeword (carrier symbol) to another valid Wi-Fi 802.11b codeword (backscatter symbol). For instance, shifting the phase of the carrier symbol by $\pi/2$ and $\pi$ modulates data 01 and 10, respectively. FreeRider [57] further extends codeword translation to Wi-Fi 802.11g/n, ZigBee, and Bluetooth signals. However, the data rate that can be achieved by such symbol-level codeword translation is usually limited to a maximum of a few hundred Kbps. Furthermore, since the tag data is encoded by the translation of carrier symbols, the receiver has to collect both the carrier signal and backscatter signal and conduct a symbol-level comparison for tag data decoding, which cannot be accomplished on a single commodity Wi-Fi receiver. WiTAG [22] encodes tag data by selectively corrupting the Wi-Fi carrier signal's subframes, but only works with frame aggregation mode.

**Frequency modulation**. The backscatter signals will collide with carrier signals when they are transmitting in the same frequency band. To avoid such interference, the third group of works proposes to shift the backscatter signal to a non-overlapping frequency band. FS-Backscatter [58] proposes an ultra-low power ring-oscillator to shift the carrier signal by 20 MHz. PLoRa [46] generates two different baseband signals to shift the backscatter signal so that these two shifted signals are merged into a valid LoRa signal at an overlapping band. To avoid the reliance on the oscillator, SHIFT [47] proposes to offload the frequency shifting signal generation to the carrier source by using twin carrier tones on different frequency bands.

Table 1: Summary of existing backscatter systems.

| System Name | Supported Modulation | Supported Protocol | Technology |
|---|---|---|---|
| Passive RFID [34, 53] | OOK | RFID | Modulate signal amplitude by varying resistive impedance |
| Amb. Backscatter [42] | OOK | N/A | |
| Turbocharging [45] | OOK | N/A | |
| WiFi Backscatter [38] | OOK | N/A | |
| mmTag [44] | OOK | N/A | |
| Passive WiFi [39] | BPSK | WiFi 802.11b | Phase shifting using delay lines (Interscatter changes phase by switching between four impedance states) |
| Interscatter [35] | DBPSK/DQPSK/OQPSK | WiFi 802.11b ZigBee | |
| HitchHike [56] | BPSK | WiFi 802.11b | |
| BackFi [23] | BPSK/16PSK | WiFi 802.11b/g | |
| FreeRider [57] | BPSK/16PSK/OQPSK | WiFi 802.11g/n ZigBee | |
| WiTAG [22] | BPSK | WiFi 802.11n/ac | |
| FS-Backscatter [58] | OOK | N/A | Frequency shifting using RF reflectors |
| FM-Backscatter [52] | FSK | FM Radio | |
| BLE-Backscatter [27] | FSK | Bluetooth | |
| PLoRa [46] | CSS | LoRa | |
| LoRa Backscatter [49] | CSS | LoRa | Frequency synthesis |
| **RF-Transformer** | CSS BPSK/16QAM/ OQPSK/ FSK | LoRa WiFi 802.11b/g/n/ac ZigBee Bluetooth | Modulate signal amplitude and phase by varying both resistive and reactive impedance |

**Difference with existing backscatter systems**. Our design differs from existing works in the following three aspects. First, they adopt different modulation technology. RF-Transformer is a flexible and programmable backscatter design, but most of existing works are tailored to a specific wireless communication protocol. A comprehensive comparison of existing backscatter systems is shown in Table 1. Second, the backscatter signals generated by our design and other works have different compatibility with standard protocols. RF-Transformer can generate protocol-compliant backscatter signals that are readily decodable on commodity wireless devices. Other works based on codeword translation require either extra hardware or the modification to firmware (for PHY-layer information retrieval). Third, the achievable throughput of these works is different. RF-Transformer supports chip-level modulation and can achieve a throughput of up to 23.8 Mbps. Whereas, most of other works modulate data at the symbol-level and thus suffer from low throughput.

The most relevant works to our design are those on generating protocol-compliant backscatter signals. Multiscatter [28] reflects different excitation signals and then modulates data on parts of the carrier symbols to convey data. Although Multiscatter supports multiprotocol backscatter for the same tag, its versatility essentially depends on the various excitation signals. It sacrifices a part of the payload in the carrier packet as the reference symbol, thus limiting the throughput. Interscatter [35] transforms the Bluetooth signal into a tone by manipulating its payload. The backscatter tag then modulates this tone into standard Wi-Fi 802.11b or ZigBee signals. However, Interscatter lacks flexibility due to their adoption of fixed load impedance (*i.e.*, can only generate Wi-Fi 802.11b and ZigBee, but not the others). Other IQ modulation works [25] support low-power LoRa backscatter link or provide high-order modulation for the millimeter-wave backscatter link [43]. Similar to these IQ modulation works, RF-Transformer explores the reflection coefficient to generate backscatter signals. Not limited to LoRa or millimeter-wave signals, RF-Transformer supports amplitude, phase, and frequency modulation through a programmable reflection coefficient design.

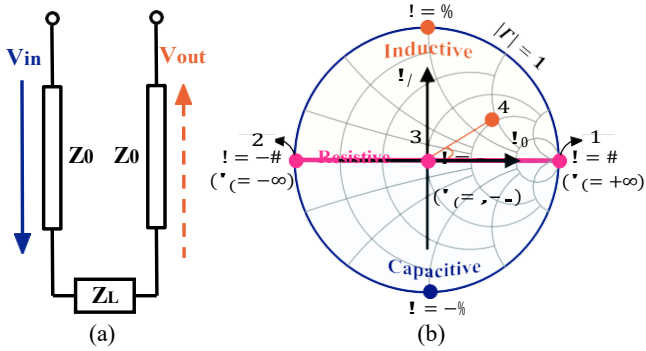

**Figure 2: Illustration of reflection coefficient.** (a) Transmission line diagram. (b) Smith chart shows system's impedance.

## 3 A UNIFIED BACKSCATTER RADIO HARDWARE ABSTRACTION

In this section, we first introduce RF-Transformer's design principle (§3.1). We then describe technical challenges (§3.2) and our proposed solutions (§3.3).

### 3.1 Basic Idea

RF-Transformer is inspired by ON-OFF Keying (OOK) modulation [26] in RFID systems. To transmit data from the tag to the reader, a load resistor connected in parallel with the RFID's antenna is switched on and off in time to change the amplitude (i.e., power) of reflected signals. This process can be better explained using transmission-line theory [16]. As shown in Figure 2(a), a proportion of incident voltage $V_{in}$ is reflected by the transmission-line and returned as $V_{out}$. We further define $\Gamma$ as the *reflection coefficient* of this power line, which is determined by the degree of mismatch between the source impedance $Z_0$[3] and the load impedance $Z_L$ [10].

$$\Gamma = \frac{V_{out}}{V_{in}} = \frac{Z_L - Z_0}{Z_L + Z_0} = \Gamma_r + j\Gamma_i \quad (1)$$

Based on the above equation, RFID varies its load impedance $Z_L$ to alter the reflection coefficient, thereby influencing the power of reflected signals. For instance, it alternates between two states: reflecting ($|\Gamma| = 1$) or absorbing ($|\Gamma| = 0$) to encode a bit "1" or "0", respectively. In practice, however, the reflection coefficient $\Gamma$ is a complex value[4] because RF systems use both resistive (e.g., resistors) and reactive components (e.g., capacitors and inductors) to balance their load impedance. As a result, both the *amplitude* and *phase* of reflected signals will be influenced by the reflection coefficient. It is understandable that any type of wireless signal can be uniquely characterized by its amplitude, phase, and frequency. Considering that the frequency variation can be derived by the phase variation since that phase is the integral of frequency, it is thus possible to generate different backscatter signals by altering the load impedance $Z_L$ connected to the backscatter radio's antenna.

### 3.2 Load Impedance Modulation and Technical Challenges

Smith chart [15] shown in Figure 2(b) is an effective tool to analyze the reflection coefficient of an RF system. It characterizes how the load impedance affects reflected signals in terms of phase and amplitude. For instance, when source impedance $Z_0$ perfectly matches load impedance $Z_L$, i.e., $Z_L = Z_0$, the reflection coefficient $\Gamma$ becomes zero, indicating the incident signal gets absorbed by the antenna without reflection (i.e., the center point "O" in the Smith chart). In contrast, when the source impedance $Z_0$ and the load impedance $Z_L$ are totally mismatched, i.e., $Z_L \ne Z_0$, the reflection coefficient $\Gamma$ becomes "+1". In this case, the incident signal will be totally reflected by the radio antenna (i.e., the point "A" in Smith chart)[5]. $Z_0$ and $Z_L$ are not strictly matched in all remaining cases. For instance, suppose $Z_L$ and $Z_0$ are $(100 + j50)\,\Omega$ and $50\Omega$, respectively. The reflection coefficient $\Gamma$ thus equals "0.4+0.2j". In this case, the amplitude of reflected signals will drop to $|\Gamma| = \sqrt{0.4^2 + 0.2^2} = 0.45$ of the incident signal. Likewise, the phase will shift by $\angle\Gamma = \arctan(0.2/0.4) = 26.6°$ with respect to the incident signal (the point "P" in Smith chart).

While the idea of load impedance modulation is compelling, practicing this idea is still challenging. To modulate both the amplitude and phase of reflected signals, we have to jointly consider the resistive (i.e., resistors) and reactive (i.e., capacitors and inductors) load connected to the radio antenna. An intuitive solution would be building a lookup table (i.e., resistance & reactance ↔ amplitude & phase) through the guidance of the Smith chart. However, to make this solution work, we have to design a dynamic matching network that allows backscatter radio to change both its resistive and reactive load using a switching circuit. Albeit low power, such design is neither scalable nor forward-compatible because wireless signals abiding by different protocols differ drastically in their amplitude and phase settings. It is difficult, if not impossible, to enumerate all load impedance through the combination of resistors, capacitors, and inductors. Besides, it also adds weight, cost, and form factor to backscatter radios, blocking their wide deployment.

### 3.3 Programming the Load Impedance

To program the load impedance flexibly, we leverage the insight that any RF signal can be represented as a complex value on a constellation diagram based on Euler's formula.

$$\begin{aligned}s(t) &= A(t)e^{j\phi(t)} \\ &= A(t)[\cos(\phi(t)) + j\sin(\phi(t))] = I(t) + jQ(t)\end{aligned} \quad (2)$$

where $A(t)$ is the signal amplitude; $\phi(t)$ is the signal phase. $I(t)$ and $Q(t)$ are in-phase and quadrature components, respectively. The signal amplitude and phase can be further represented by:

$$\begin{aligned}A(t) &= \sqrt{I^2(t) + Q^2(t)} \\ \phi(t) &= \arctan(\frac{Q(t)}{I(t)}) = \arctan(\frac{|Q(t)|}{|I(t)|}) + \theta(t)\end{aligned} \quad (3)$$

where $\theta(t)$ is the phase of the carrier signal. The above equation indicates that one can generate any backscatter waveforms by altering only the *amplitude* of their in-phase and quadrature components. Accordingly, we now transform the sophisticated resistive and reactive dual-load control into resistive-only load control.

---
[3] Any source of power comes with an impedance, denoted as $Z_0$.
[4] Where $\Gamma_r$ and $\Gamma_i$ represent the real and imaginary part.

[5] When $Z_L = \infty$, the amplitude of reflected signals is the same as that of the carrier signal but opposites in phase (denoted as "B" in Smith chart).

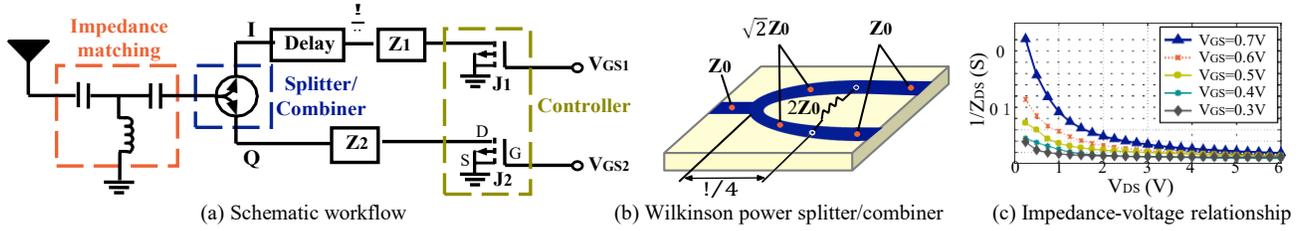

Figure 3: The circuit design of RF-Transformer. (a) Schematic workflow. (b) Wilkinson power splitter/combiner. (c) Impedance-voltage relationship of a typical transistor ATF-54143 [12].

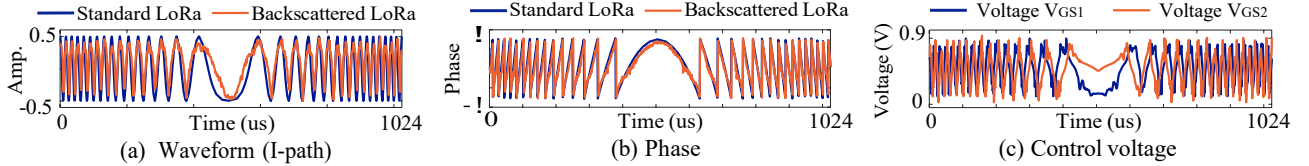

Figure 4: LoRa symbol synthesis. (a) Standard LoRa signal and backscattered LoRa signal. (b) Standard LoRa phase and backscattered LoRa phase. (c) Control voltage $V_{GS1}$ and $V_{GS2}$ to synthesize the desired LoRa backscatter symbol.

**Build the circuit**. Based on the above observation, we build an ultra-low power circuit to program the amplitude of in-phase $I(t)$ and quadrature component $Q(t)$ individually by changing the resistive load connected to each part.

Figure 3(a) shows the schematic of this circuit. The incident signal (*i.e.*, carrier signal) first goes through an impedance matching circuit to ensure most of their energy gets absorbed by the antenna. We get two copies of the incident signal by passing it through a passive Wilkinson power splitter [18]. These two signal copies propagate along two separate paths. We add a delay line to one of the paths to shift the phase of the signal propagating along this path by $\frac{\pi}{4}$. When this signal goes in and then out from the same delay line, it will experience $\frac{\pi}{2}$ delay with respect to the signal propagation along another path. Consequently, these two signals form the in-phase $I(t)$ and quadrature $Q(t)$ component of backscatter signals. We optimize the length, width, and material of the microstrip line through simulation to ensure that the signal power on two paths is symmetric and of equal strength (Figure 3(b)).

To modulate the signal power along these two paths, we connect each path to a MOSFET transistor consisting of a gate (G), a source (S), and a drain (D). The resistive load impedance of this transistor can be altered by varying its bias voltage ($V_{GS}$). Figure 3(c) shows the resistive load impedance of this transistor (ATF-54143 [12]) as a function of bias voltage. Evidently, as we increase the bias voltage, the resistive load impedance of this transistor drops gradually, which further changes the amplitude of reflected signal. It thus allows the backscatter radio to generate different waveforms by changing the amplitude of in-phase $I(t)$ and quadratic component $Q(t)$.

Although RF-Transformer shares the same design principle with the traditional QAM transceiver [13], they separate and modulate in-phase and quadrature components in different ways. The traditional QAM transceiver uses high-precision oscillator and DAC to generate baseband in-phase and quadrature components, respectively. In contrast, RF-Transformer neither generates baseband nor RF carrier signals. It instead modulates the in-phase and quadrature components of an ambient carrier signal by altering the reflection coefficient.

### 3.4 Extending to Frequency Modulation

The above load impedance modulation circuit allows RF-Transformer to synthesize any type of amplitude-modulated signals or phase-modulated signals, or both. We next explain synthesizing frequency-modulated signals using this design.

We take LoRa as an example. LoRa adopts Chirp Spread Spectrum (CSS) to modulate data [40]. Each LoRa symbol is represented by a chirp whose frequency changes linearly over time, represented by $f(t) = F_0 + kt$, where $F_0$ is the initial frequency offset; $k$ is the frequency changing rate, which is known in advance. It is understandable that the instantaneous frequency $f(t)$ of a LoRa symbol can be derived by the phase changing rate using the equation $f(t) = \frac{1}{2\pi} \frac{d\Phi(t)}{dt}$. Accordingly, one can emulate the frequency variation of a LoRa symbol by manipulating its phase variation. The instant phase of a LoRa symbol at time $t$ can be derived by:

$$\Phi(t) = 2\pi \int_0^t f(\tau)\, d\tau = 2\pi \int_0^t (F_0 + k\tau)\, d\tau \\ = 2\pi (F_0 t + \tfrac{1}{2} kt^2) = \Phi_0 + \pi k t^2 \quad (4)$$

where $\Phi_0$ is the initial phase of this chirp symbol. Based on Eq. 4, RF-Transformer can derive the phase value at each time point, and further leverage load impedance modulation to synthesize the desired chirp symbols.

We experimentally demonstrate the feasibility of such frequency modulation. Specifically, we use a USRP N210 software-defined radio to generate a sinusoidal carrier signal with 1 MHz bandwidth in 2.45 GHz frequency band. The backscatter radio modulates this carrier signal into a standard LoRa symbol "100011". The bandwidth and spreading factor of this LoRa symbol are set to 125 KHz and 7, respectively. Another USRP N210 is employed to receive this backscatter signal. The receiver's sampling rate is set to 250 KHz.

Figure 4(a) shows the received backscatter waveform. We extract the phase readings and plot them in Figure 4(b). The bias voltage applied to $V_{GS1}$ and $V_{GS1}$ at each time point is shown in Figure 4(c). We observe that the synthesized LoRa symbol resembles the standard LoRa symbol in terms of both waveform and phase pattern. The receiver can successfully demodulate this backscattered LoRa signal through dechirping, confirming our analysis. We further evaluate backscattered LoRa signal in various SNR conditions in Section 7.1.

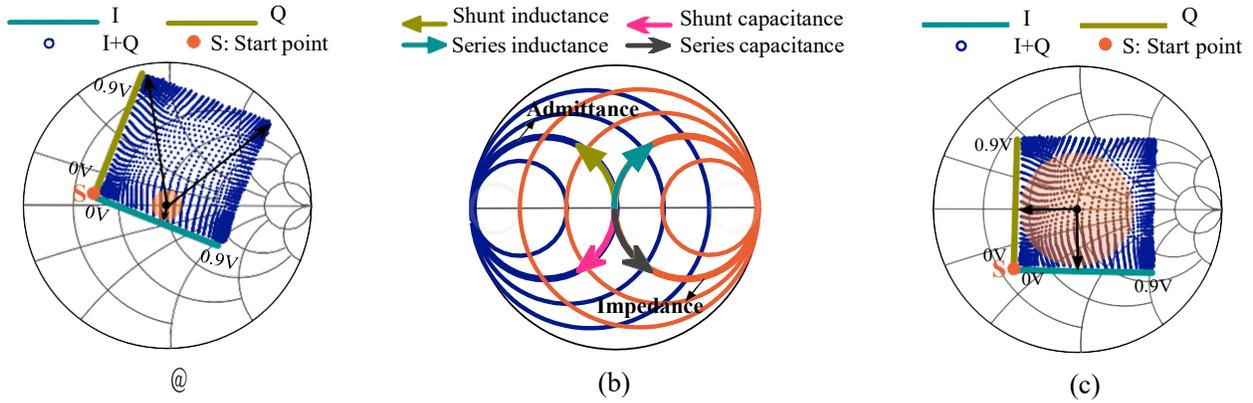

**Figure 5: Modulation space of RF-Transformer.** (a): Modulation space of RF-Transformer before optimization. (b): Optimize modulation space through impedance correction. (c): Modulation space of RF-Transformer after optimization.

## 4 RF-TRANSFORMER: PRACTICAL DESIGN CONSIDERATIONS

Although the above design can generate any type of wireless signal, we have to address the following two practical issues to fully unleash its potential.

### 4.1 Optimizing the Modulation Space

In practice, all hardware components that are indirectly connected to the radio antenna also bring a fixed load, which negatively impacts the modulation space of backscatter signals. To better understand this issue, we sweep the bias voltage of both two transistors from 0 V to 0.9 V with a step of 1 mV. In each bias voltage setting, we plot the equivalent impedance of the backscatter radio in the Smith chart, shown in Figure 5(a). These impedance points form two borderlines, corresponding to two cases explained below:

- Varying the bias voltage of the $I$ path while fixing the bias voltage of $Q$ path to 0 V (dark green line).
- Varying the bias voltage of the $Q$ path while fixing the bias voltage of $I$ path to 0 V (light green line).

The dashed points in blue represent a set of load impedance that can be generated through the change of two transistors connected to the I-Q paths. To facilitate the presentation, we term the coverage area of these dashed points as the *modulation space of this backscatter radio*. The intersection point $S$ of these two borderlines denotes the fixed load impedance introduced by the hardware board.

This fixed load impedance translates and rotates the modulation space on the Smith chart. To guarantee the continuity and completeness of amplitude settings over the entire phase space (*i.e.*, $[0, 2\pi]$), the effective modulation space of this radio will shrink to a circular area (*i.e.*, the area in orange), indicating that the backscattered signals tend to be weak and their communicate range will suffer accordingly.

To retain a large effective modulation space, we design a matching circuit and add it between the radio antenna and power splitter to correct the translation and rotation offset caused by the radio board's fixed load impedance. One thing worth mentioning is our matching circuit design differs from the conventional impedance matching for RF systems in their optimization goals: the impedance matching aims to minimize reflection coefficient whereas our matching circuit aims to maximize the effective modulation space.

The proposed matching circuit corrects the modulation space by shunt inductance/capacitance or series inductance/capacitance to change the impedance of the circuit. The difference between the shunt inductance/capacitance and series inductance/capacitance determines that the impedance of the circuit presents capacitive or inductive characteristics, as shown in Figure 5(b). The circuit design principles are as follows:

- Adding *inductors* in series to the circuit will rotate the modulation space clockwise with respect to the impedance circle, while adding *capacitors* in series to the circuit will rotate the modulation space counterclockwise, as shown in Figure 5(b).
- Adding *shunt capacitors* to the circuit will rotate the modulation space clockwise with respect to the admittance circle, whereas adding *shunt inductors* will rotate the modulation space counterclockwise.

Figure 5(c) shows the optimized modulation space after adding the proposed matching circuit. Compared to the initial modulation space shown in Figure 5(a), we can see the effective modulation space (the area in orange) of RF-Transformer becomes significantly larger. This allows RF-Transformer to generate different types of backscatter signals without sacrificing too much of communication range.

### 4.2 Configuring the Carrier's Baseband

RF-Transformer synthesizes different types of backscatter signals by modulating the carrier signal – a sinusoidal tone. The baseband frequency $F_B$ of this tone, the modulation frequency of backscatter tag $F_M$, and the sampling rate $F_S$ of the receiver are critical to the success of signal demodulation and cannot be set arbitrarily. It's worth noting that the baseband frequency of the carrier signal is irrelevant with carrier frequency offset (CFO). To better understand this issue, we program RF-Transformer to synthesize symbol 00 01 11 10 repeatedly based on 4-QAM modulation (Figure 6(a)). We then plot the received backscatter signal on the I-Q plane.

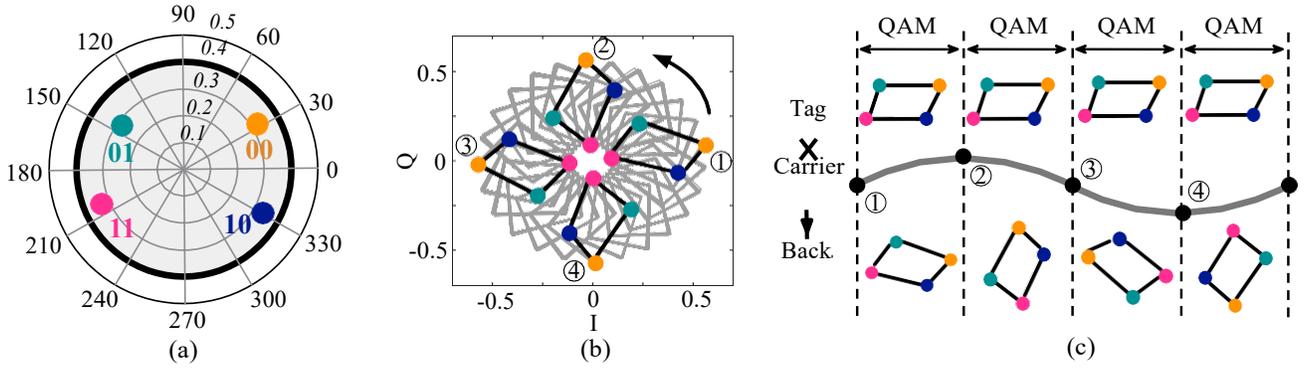

**Figure 6: The received backscatter signal when $F_B < F_M$.** (a) 4-QAM modulation. (b) Each four consecutive signal samples in the I-Q plane form a quadrilateral, corresponding to four different QAM symbols, and these quadrilaterals spin over time. (c) The initial phase of the baseband carrier is different for each round of QAM modulation, resulting in the spinning of QAM points.

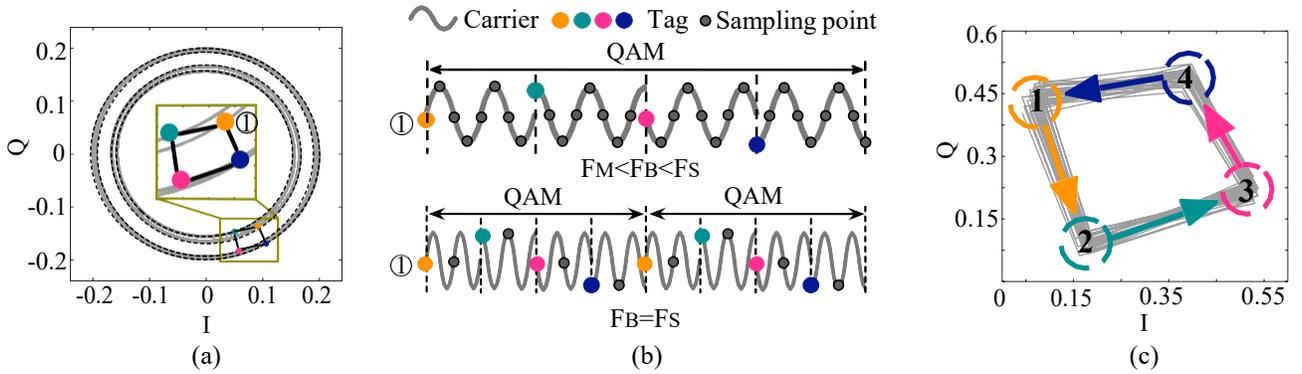

**Figure 7: Signal demodulation in different cases.** (a) Signal samples on the I-Q plane when $F_M < F_B < F_S$ (b) Analysis of time-domain waveform. (c) Signal samples on the I-Q plane when $F_B = F_S$.

We first show that the baseband frequency of carrier signals should be larger than the modulation frequency of the backscatter tag (i.e., $F_B > F_M$). Suppose $F_B < F_M$, without loss of generality, we set $F_B$ and $F_M$ to 40 KHz and 1 MHz respectively. The receiver samples the backscattered signals at 2 MHz. We plot the received backscatter signal samples in Figure 6(b). From this figure we observe each four consecutive signal samples in the I-Q plane form a quadrilateral, corresponding to four different QAM symbols. These quadrilaterals spin over time, hence the received signals are not decodable on the receiver. The reason behind this phenomenon can be better explained by Figure 6(c). Recall that $F_M$ is larger than $F_B$, hence the backscatter radio will synthesize multiple groups of QAM symbols within each carrier symbol ($[0,2\pi]$). Since the phase of carrier signal changes over time within each carrier symbol, the initial phase of each QAM symbol will change accordingly. Therefore, the received backscatter symbols will spin over time on the I-Q plane.

We next demonstrate that the receiver's sampling frequency should be no larger than the carrier's baseband frequency (i.e., $F_S \leq F_B$). Suppose $F_B < F_S$, without loss generality, we set $F_B$, $F_M$, and $F_S$ to 2 MHz, 250 KHz, and 24 MHz, respectively. i.e., $F_M < F_B < F_S$. We observe that the received signal samples span the entire phase space from 0 to $2\pi$, as shown in Figure 7(a). These signal samples form four concentric circles in the I-Q plane due to the amplitude variation, which negatively affects the signal demodulation. We use the top figure in Figure 7(b) to explain this issue. Recall that $F_B$ is larger than $F_M$, hence the backscatter tag modulates one QAM symbol every few carrier symbols. However, since we suppose the sampling fequency of the receiver is larger than the basedband frequency of the carrier signal (i.e., $F_B < F_S$), we expect to see many signal points (the grey points in the top figure of Figure 7(b)) whose phase and amplitude both change over time.

Thus we have: $F_B \geq F_S \geq 2F_M$. On the one hand, the receiver undersamples the carrier signal so that the carrier signal cannot be recovered and will not interfere the demodulation of the backsactter signal. On the other hand, based on the Nyquist–Shannon sampling theorem, the receiver's sampling frequency is two times higher than the bandwidth of backscatter signals, which allows the receiver to demodulate the backscatter signals successfully. As an example, we set $F_B$, $F_M$, and $F_S$ to 1 MHz, 500 KHz, and 1 MHz respectively and repeat the above experiment. Figure 7(c) shows the received signal samples on the I-Q plane. Evidently, these signal samples form four clusters, corresponding to four different QAM symbols. The stability of these symbol clusters indicates that the receiver can reliably decode each backscatter symbol.

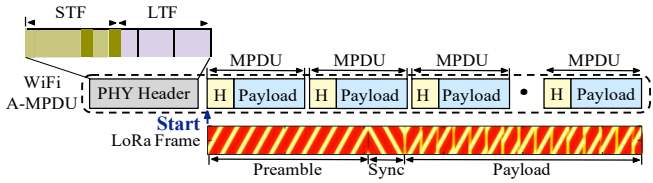

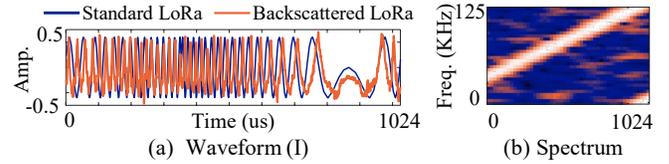

**Figure 8: RF-Transformer takes WiFi signal as the carrier to synthesize LoRa frame.** The short training field (STF) of WiFi is used for packet detection and synchronization. WiFi 802.11n/ac frame aggregation mechanism extends the WiFi frame length for synthesis of LoRa backscatter frame.

**Figure 9: RF-Transformer synthesizes LoRa backscatter signal by taking the ambient WiFi signal as the carrier.** (a) Standard LoRa signal and backscattered LoRa signal. (b) The spectrum of backscattered LoRa signal.

## 5 RF-TRANSFORMER: CROSS-TECHNOLOGY BACKSCATTER

We demonstrate that RF-Transformer can also support cross technology [29, 37, 41, 54] backscatter. As a proof of concept, we seek to modulate Wi-Fi signals into LoRa packets by changing the bias voltage of two transistors on a backscatter tag. Enabling cross-technology backscatter faces the following challenges.

First, unlike continuous sinusoidal tone, Wi-Fi traffic is intermittent. It coexists with other types of wireless traffic (*e.g.*, Bluetooth, ZigBee) on the same unlicensed band. Consequently, the backscatter tag should be able to distinguish the Wi-Fi signal from the others and further synchronize with Wi-Fi symbols for fine-grained modulation. To achieve this goal, we program the backscatter tag to sample the incident signals and perform cross-correlation with the pre-stored Short Training Field (STF) of the Wi-Fi packet.

Second, the difference in frame length between Wi-Fi (200 $\mu$s) and LoRa (1 ms) makes it impossible to generate a complete LoRa frame on top of a single Wi-Fi frame. Modulating multiple consecutive Wi-Fi frames into a LoRa frame may run into issues since the inter-frame spacing between Wi-Fi frames varies with the payload [48]. Hence, we are expected to see fragments in the modulated LoRa symbols. Our solution is to take advantage of the frame aggregation feature [22, 24, 36] on existing Wi-Fi systems to accommodate LoRa symbol modulation. To improve MAC efficiency and channel utilization, existing Wi-Fi transmitters combine multiple Wi-Fi frames into a single transmission unit (*i.e.*, MAC Protocol Data Unit Aggregation, A-MPDU) with a single physical header, as shown in Figure 8. The aggregated frame length of A-MPDU (64 KB) can reach up to 72 ms, which is enough to modulate 72 LoRa symbols.

Third, both the amplitude and phase of Wi-Fi signals may change abruptly. Notice that the absolute phase value of the backscatter signal depends on the carrier's phase (Equation 3). Accordingly, an abrupt change of Wi-Fi carrier's phase value will mess up the phase modulation of backscatter signals. Like Interscatter [35], we leverage the signal emulation techniques to generate a single-tone by manipulating the payload of Wi-Fi frames [41]. This ensures RF-Transformer modulates backscatter symbol on top of the carrier symbol with the same initial phase.

Figure 9(a) compares the waveform of a standard LoRa chirp symbol "1111100" and our synthesized LoRa symbol on top of Wi-Fi signals. Evidently, these two waveforms share very similar patterns. We further plot the spectrum of this backscatter symbol in Figure 9(b). The frequency of this chirp symbol changes continuously over time, resembling the standard LoRa chirp.

## 6 IMPLEMENTATION

We describe system implementation in this section.

### 6.1 Backscatter Tag

The RF-Transformer tag consists of a RF front-end and a microprocessor unit. We implement the RF front-end on a 65 $mm\times 37$ $mm$ one-layer PCB using commercial off-the-shelf analog components. Figure 10(a) shows the hardware prototype. The RF front-end is completely passive. We connect it to an omnidirectional antenna [3] with 3 dBi gain. The impedance matching circuit consists of two inductors connected in parallel and two capacitors connected in series. Two RF MOSFETs ATF-54143 [12] terminate the I-path and Q-path. We change the impedance of the MOSFET transistor by controlling the bias voltage, so as to synthesize the desired backscatter signal. The switching rate of the MOSFET transistor is equal to the data rate of the backscatter signal. Since the impedance of the MOSFET transistor varies non-linearly with the input voltage as shown in Figure 3(c), we need to conduct additional tuning and calibration before controlling the MOSFET's impedance. To avoid the in-band interference, RF-Transformer uses a RF switch ADG901 [14] to shift the backscatter signal to another frequency band 25 MHz away from the carrier's band.

The microprocessor unit shown in Figure 10(b) consists of a low-power Xilinx Artix-7 FPGA [19] and a 14-bit DAC AD9767 [1]. They work together to control two MOSFET transistors. The Artix-7 FPGA has an oscillator of 50 MHz that enables the maximum modulation rate of Wi-Fi 20 MHz. The 14-bit DAC AD9767 achieves the voltage accuracy of 0.3 mV given the reference voltage 5 V. We use a Vector Network Analyser (VNA) CEYEAR 3672A [17] to measure the voltage-impedance response of the transistor and store the results in FPGA. Figure 10(c) shows the setup.

### 6.2 Transmitter and Receiver

**Transmitter.** We use a Software-Defined Radio (SDR) platform USRP N210 to transmit continuous sinusoidal signals. A Mini-Circuits RF amplifier [20] is used to boost the signal power up to 30 dBm[6]. The amplified carrier signals are sent out using an omnidirectional antenna with 3 dBi gain [3]. Besides SDR platform, there are other two methods that can provide low-cost carrier. The first method is to build a low-cost transceiver that generates the sinusoidal carrier using Direct Digital Synthesis (DDS) [50]. The second method uses commercial devices such as a smartphone to generate the sinusoidal carrier by manipulating the payload of their transmitted Bluetooth or Wi-Fi packet [35].

---
[6] According to Federal Communications Commission (FCC) regulation, the transmission power of a frequency radio should be below 32 dBm [5].

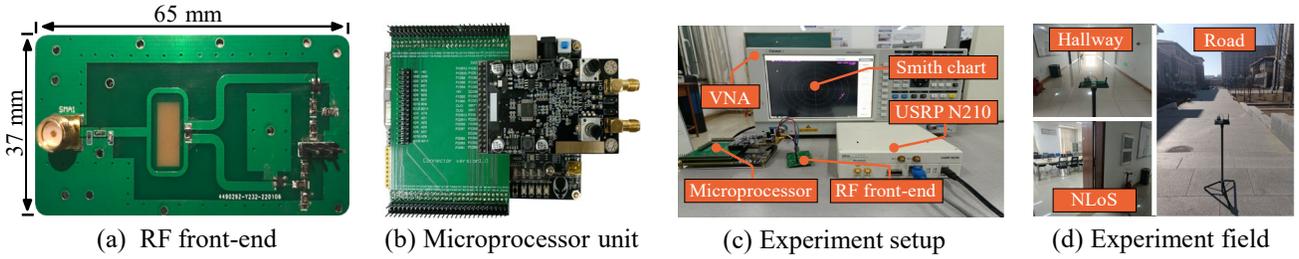

(a) RF front-end  (b) Microprocessor unit  (c) Experiment setup  (d) Experiment field

Figure 10: RF-Transformer prototype and experiment setups.

**Receiver.** We use an USRP N210 equipped with the same type of antenna as the receiver. The receiver runs open-sourced gr-ieee802-11 [7], gr-ieee802-15-4 [9], gr-bluetooth [8], and gr-lora [11] to receive and decode the backscattered Wi-Fi, ZigBee, Bluetooth, and LoRa signals.

### 6.3 Downlink and Node Coordination

The carrier transmitter leverages OOK-modulated query message to inform the RF-Transformer tag when to backscatter, what type of excitation carrier, and what type of signal to synthesize. The RF-Transformer tag listens to the channel and down converts the OOK-modulated carrier signal to the baseband by using an envelope detector. Then the low-power FPGA decodes the query message and further generates the standard-compliant packet. The envelope detector is a passive component (including *i.e.* inductance, capacitors, and diodes), hence the energy consumption of this downlink demodulation circuit mainly comes from the FPGA. Our FPGA works on the low sampling rate mode and the detection alogorithm is simple. The total power consumption of the downlink demodulation circuit is around 85 $\mu$W. When a network consists of multiple RF-Transformer tags, the carrier transmitter allocates different time slots to different RF-Transformer tags through query messages. The RF-Transformer tags abide by the Time Division Multiple Access (TDMA) protocol to transmit backscatter signals.

### 6.4 ASIC Simulation

We simulate the Application Specific Integrated Circuit (ASIC) of RF-Transformer based on the TSMC 65-nm CMOS process. The active area of on-chip Integrated Circuits (IC) is 0.535 $mm^2$. The ASIC simulation result shown in Table 2 reports that the power consumption of RF-Transformer to synthesize LoRa, BLE, ZigBee, Wi-Fi 802.11b, and Wi-Fi 802.11g/n/ac signal is 47.8 $\mu$W, 80.1 $\mu$W, 89.2 $\mu$W, 79.0 $\mu$W, and 371.2 $\mu$W, respectively. Among these hardware components, the most power-hungry parts are FPGA and DAC, which account for 20.96%–25.52% and 69.25%–73.95% of the total power consumption respectively for synthesizing different backscatter signals.

## 7 EVALUATION

In this section, we first present the field studies on synthesizing different types of backscatter signals (§7.1). We then conduct micro-benchmarks to understand the key factors impacting the system performance (§7.2). Finally, we measure the power consumption of RF-Transformer tag (§7.3) on generating different wireless signals. A real-world case study on detecting foot traffic density in hallways follows (§7.5).

**Metric**. We adopt *BER*, *throughput*, and *backscatter range* as the key metrics to assess RF-Transformer's performance.

- **BER** refers to the ratio of error bits to the total number of bits received by RF-Transformer.
- **Throughput** measures the amount of received data correctly decoded by RF-Transformer within one second.
- **Backscatter distance** refers to the maximum distance between the tag and the receiver when the BER is maintained below 1%.

### 7.1 Field Studies

We control the RF-Transformer tag to synthesize multiple protocol-compliant backscatter signals, including LoRa, Wi-Fi 802.11b, Wi-Fi 802.11g/n/ac, ZigBee, and Bluetooth. We conduct end-to-end studies to assess RF-Transformer's performance for different backscatter links.

**LoRa backscatter link**. We control the RF-Transformer tag to generate LoRa backscatter signals. We place the RF-Transformer tag 20 cm away from the transmitter, and move the receiver 100 m, 200 m, 400 m, 600 m, 800 m away from the RF-Transformer tag. Under each distance setting, we vary the spreading factor (SF) and bandwidth (BW) to assess the LoRa backscatter link's performance.

First, we set BW to 250 KHz, vary SF from 7 to 12, and measure BER and throughput. As shown in Figure 11, we observe that the BER declines with the increasing SF. For instance, the BER under the highest SF setting (*i.e.*, SF=12) is 4.7–35$\times$ lower than that under the lowest SF setting (*i.e.*, SF=7) across all tag-to-receiver distances. Given the BER less than 1%, the achievable backscatter distance declines from 643.5 m to 245.8 m when the SF declines from 12 to 7. This is expected since a higher SF enhances the anti-noise capability of LoRa signals, and thus the backscatter distance grows. On the other hand, the symbol time grows with the increasing SF, resulting in lower throughput. We observe that when the SF

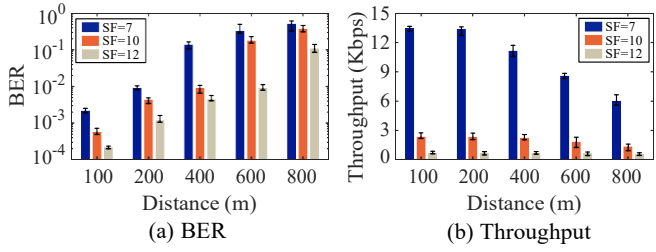

Figure 11: Performance of LoRa backscatter link in different SF settings.

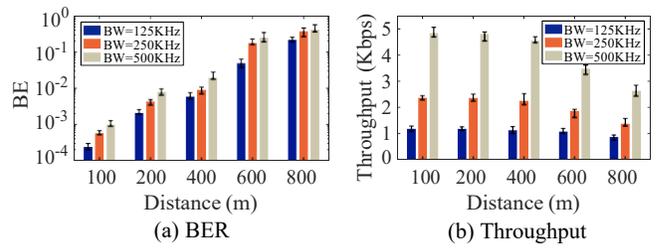

Figure 12: Performance of LoRa backscatter link in different BW settings.

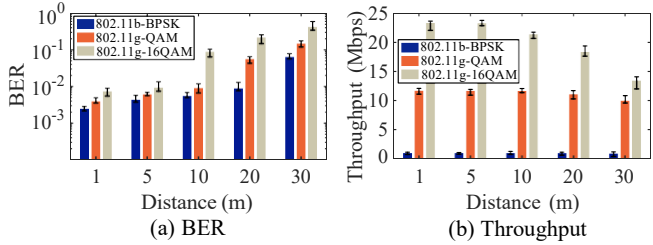

Figure 13: Performance of Wi-Fi backscatter link.

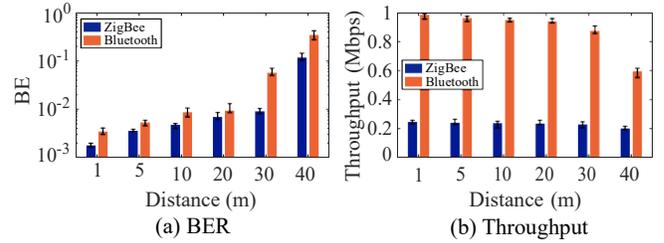

Figure 14: Performance of ZigBee and Bluetooth backscatter links.

declines from 12 to 7, the throughput grows 10.1–18.6× across all tag-to-receiver distances.

Second, we set SF to 10 and vary the BW from 125 KHz to 500 KHz. The evaluation result is shown in Figure 12. We observe that the BER and the throughput both grow with the BW. Specifically, given the tag-to-receiver distance of 200 m, the BER grows from 0.0021 to 0.008 as we increase the BW from 125 KHz to 500 KHz. On the other hand, since the LoRa symbol time is inversely proportional to the BW, we observe the throughput drops 3.1–4.2× as we decrease the BW from 500 KHz to 125 KHz across all tag-to-receiver distances.

**Wi-Fi backscatter link**. We control the RF-Transformer tag to generate Wi-Fi signals, including Wi-Fi 802.11b with modulation of BPSK, Wi-Fi 802.11g/n/ac with modulation of QAM and 16 QAM, respectively. Although Wi-Fi 802.11g/n/ac adopts Orthogonal Frequency Division Multiplexing (OFDM) technology, the Wi-Fi OFDM signal combined by multiple subcarriers can still be uniquely characterized by amplitude, phase and frequency. Therefore, RF-Transformer supports the Wi-Fi OFDM backscatter link. We place the RF-Transformer tag 20 cm away from the transmitter and move the receiver 1 m, 5 m, 10 m, 20 m, 30 m away from the RF-Transformer tag. Under each distance setting, we assess the performance of three different Wi-Fi backscatter links. We have two observations based on the evaluation result shown in Figure 13.

First, we observe that the BER and the throughput both grow with the increase of modulation complexity. When the Wi-Fi backscatter link varies from 802.11b BPSK to 802.11g/n/ac 16QAM, the BER grows 2.3–24.5×, and the throughput grows 14.7–23.9 across all tag-to-receiver distances. For example, given the tag-to-receiver distance of 10 m, the BER grows from 0.0056 to 0.09 and the throughput grows from 0.975 Mbps to 21.43 Mbps. In accordance with expectation, Wi-Fi 802.11b BPSK adopts a lower modulation rate and leverages DSSS mechanism to improve the robustness, thus resulting in lower BER and throughput. For Wi-Fi 802.11g/n/ac, the 16QAM modulation has more diverse phase and amplitude, thus resulting in higher BER and throughput.

Second, the BER grows but throughput declines with the increase of tag-to-receiver distances. Specifically, when the tag-to-receiver distance grows from 1 m to 30 m, the BER grows 26–57.1×, and the throughput drops 1.07–1.74× across all Wi-Fi backscatter links. Given the BER less than 1%, the achievable backscatter distance are 25.8 m, 16.5 m, and 8.4 m for Wi-Fi 802.11b BPSK, 802.11g/n/ac QAM, and 802.11g/n/ac 16QAM backscatter links, respectively.

**ZigBee and Bluetooth backscatter links**. We control the tag to generate ZigBee and Bluetooth backscatter signals. We place the RF-Transformer tag 20 cm away from the transmitter, and move the receiver 1 m, 5 m, 10 m, 20 m, 30 m, 40 m away from the RF-Transformer tag. Under each distance setting, we assess the performance of ZigBee and Bluetooth backscatter links. The evaluation result is shown in Figure 14. First, the BER of ZigBee backscatter link and Bluetooth link grow with the increase of tag-to-receiver distance. When the tag-to-receiver distance increases from 1 m to 40 m, the BER of ZigBee backscatter link grows by 66.67× from 0.0018 to 0.12, and the BER of Bluetooth backscatter link grows by 109.38× from 0.0032 to 0.35. Second, the throughput of ZigBee backscatter link and Bluetooth link drop with the increase of tag-to-receiver distance due to the deteriorating SNR. When the tag-to-receiver distance increases from 1 m to 40 m, the throughput of ZigBee backscatter link drops by 1.22× from 247.1 Kbps to 202.4 Kbps, and the throughput of Bluetooth backscatter link drops by 1.65× from 986.5 Kbps to 598.2 Kbps. Third, the BER of ZigBee backscatter link is 1.35–6.32× lower than Bluetooth backscatter link.

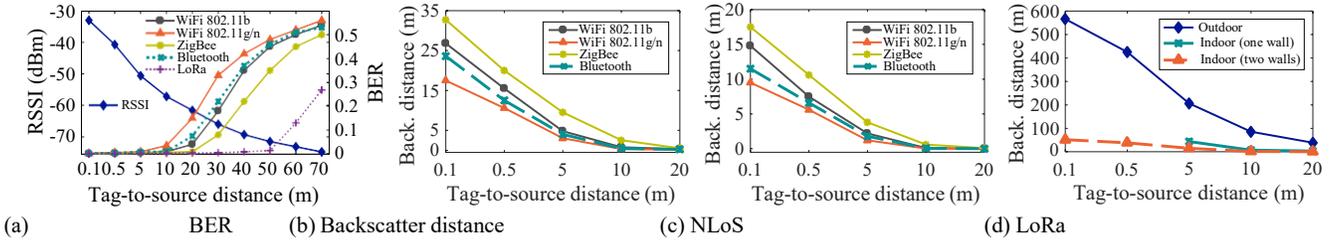

(a) BER  (b) Backscatter distance  (c) NLoS  (d) LoRa

Figure 15: RF-Transformer's performance under different tag-to-source distances.

Given the BER less than 1%, the achievable backscatter distance are 33.5 m and 21.4 m for ZigBee and Bluetooth backscatter links. ZigBee adopts DSSS mechanism which translates a 4-bit symbol into a 32-bit chip sequence. On the one hand, DSSS mechanism can resist the harmful impact caused by the signal imperfection and attenuation. On the other hand, it also leads to the decline of throughput. For example, the throughput of ZigBee backscatter link is 2.95–3.98× lower than Bluetooth backscatter link.

### 7.2 Impact of Tag-to-source Distance

In this section, we conduct experiments to evaluate the impact of tag-to-source distance on the backscatter link. We place a RF-Transformer tag 10 cm–70 m away from the transmitter and measure the BER and backscatter range of RF-Transformer.

**BER**. As shown in Figure 15(a), the Received Signal Strength (RSS) of the detectable backscatter signal decreases with the increase of the tag-to-source distance. As expected, the BER of the Wi-Fi, ZigBee, Bluetooth, and LoRa backscatter links grow. Under the same tag-to-source distance, we observe that the BER of LoRa is the lowest, ZigBee comes second, and Wi-Fi 802.11g/n is the highest. This is because the spread spectrum mechanism such as DSSS of ZigBee and CSS of LoRa improves the anti-noise ability.

**Backscatter distance**. We observe the backscatter distance drops with the grow of tag-to-source distance as shown in Figure 15(b) and Figure 15(d). When the tag-to-source distance grows from 0.1 m to 5 m, we observe the following performance: the LoRa backscatter range drops by 2.76× from 568.8 m to 206 m; the Wi-Fi 802.11b backscatter range drops by 5.58× from 26.8 m to 4.8 m; the Wi-Fi 802.11g/n backscatter range drops by 5.83× from 17.5 m to 3.1 m; the ZigBee backscatter range drops by 3.44× from 32.7 m to 9.5 m; and the Bluetooth backscatter range drops by 5.91× from 23.6 m to 4.2 m.

The backscatter range scales with the strength of backscatter signals. Due to signal attenuation, insertion loss, and energy transformation loss on the backscatter tag, the backscatter signal is orders of magnitude weaker than the active signals. This is also the reason for the limited range of the backscatter systems compared with the active technologies. Using the hardware with negative impedance [51] to reduce the energy transformation loss on the backscatter tag may be one of the solutions to improve the range, and that is our future work.

**NLoS scenario**. As shown in Figure 15(c) and Figure 15(d), the backscatter range in NLoS scenario (when backscatter signal penetrating one concrete wall) significantly drops compared with the LoS scenario shown in Figure 15(b). For example, when the tag-to-source distance is 0.1 m, the backscatter range drops by 1.81×, 1.84×, 1.86× and 2.05× for Wi-Fi 802.11b, Wi-Fi 802.11g/n, ZigBee, and Bluetooth backscatter links. Owing to the performance gain brought by the CSS, the backscatter range for the LoRa backscatter link is still longer than 100 m when penetrating one wall. It is difficult to directly measure the exact value due to the limited size of the experimental field. Whereas, we observe that the backscatter range for the LoRa backscatter link is 51.5 m after penetrating two concrete walls.

Table 2: Energy consumption of each component.

|  | Energy (uW) | | | | | ASIC (uW) | | | | |
|---|---|---|---|---|---|---|---|---|---|---|
|  | Oscillator | FPGA | DAC | RF Transistor | Total | Oscillator | FPGA | DAC | RF Transistor | Total |
| LoRa | 78.8 | 7.6 | 234.2 | 18.6 | 339.2 | 10.8 | 1.4 | 33.1 | 2.5 | 47.8 |
| Bluetooth | 134.3 | 12.5 | 475.4 | 33.5 | 655.7 | 15.7 | 1.8 | 58.4 | 4.2 | 80.1 |
| WiFi 802.11b | 140.7 | 13.8 | 527.6 | 41.7 | 723.8 | 15.4 | 1.8 | 57.3 | 4.5 | 79.0 |
| WiFi 802.11g/n/ac | 1149.4 | 86.2 | 4391.3 | 242.3 | 5869.2 | 73.8 | 7.5 | 274.5 | 15.4 | 371.2 |

Table 3: Energy consumption of active radio (LoS)

|  | Standard | | | Normalized Comparison | | |
|---|---|---|---|---|---|---|
|  | TX Power (dBm) | Range (m) | Energy (mW) | TX Power (dBm) | Range (m) | Energy (mW) |
| Bluetooth [4] | 10 | 75 | 30 | 4.7 | 23.6 | 9.5 |
| WiFi 802.11b [55] | 20 | 100 | 200 | 14.3 | 26.8 | 53.6 |

### 7.3 Power Consumption

As summarized in Table 2, RF-Transformer consumes 339.2 $\mu$W, 655.7 $\mu$W, 841.3 $\mu$W, 723.8 $\mu$W, and 5869.2 $\mu$W to synthesize LoRa, Bluetooth, ZigBee, Wi-Fi 802.11b, and Wi-Fi 802.11g/n/ac signal, respectively. Among these hardware components, the most power-hungry parts are oscillator and DAC, which account for 18.9%-23.2% and 69.1%-74.8% of the total power consumption across all synthesized signals. The power consumption of RF-Transformer is higher than other existing backscatter technologies, *i.e.* Multiscatter [28]. That is expected since RF-Transformer requires a high-speed oscillator and high-accuracy DAC to control the bias voltage of the RF transistor to generate the protocol-compliant backscatter signals. Generally, the power consumption of RF-Transformer is related to the type of the synthesized signal. The larger the bandwidth and data rate of the synthesized signal, the higher the power consumption of RF-Transformer. This is also the reason for the highest power consumption for the Wi-Fi 802.11g/n/ac backscatter link. After ASIC fabrication, the power consumption of RF-Transformer further drops to 47.8 $\mu$W–371.2 $\mu$W.

**Comparison with active radios.** Table 3 shows the power consumption of the active LoRa [2], Bluetooth [4], ZigBee [21], and WiFi [55] chip. According to the loss model of wireless signal in the free space [6], we can calculate the power consumption of these active radios in the same communication range as that of RF-Transformer. The power consumption of active LoRa, Bluetooth, ZigBee, Wi-Fi 802.11b, and Wi-Fi 802.11g/n/ac radio are 11.4 mW, 9.5 mW, 8.2 mW, 53.6 mW and 44.8 mW, respectively. Overall, the power consumption of RF-Transformer is 7.6–74.2× less than their active counterparts.

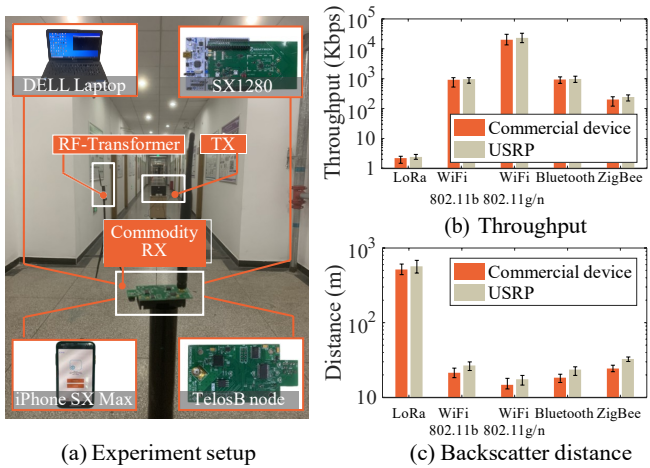

**Figure 16: RF-Transformer's performance with commercial receivers.** (a) Experiment setup. (b) Throughput. (c) Backscatter range.

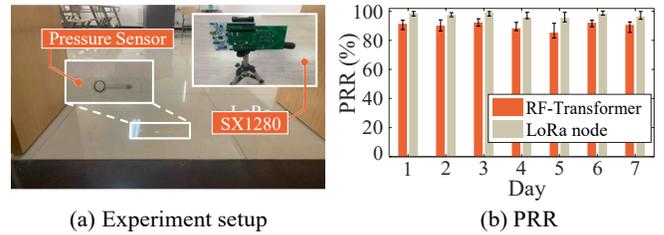

**Figure 17: Foot traffic detection application.** (a) Experiment setup. (b) PRR.

### 7.4 Compatibility with Commodity Receivers

In order to verify the compatibility of RF-Transformer with commodity receivers, we deploy a commercial LoRa node SX1280, a DELL laptop, an iPhone XS Max smartphone, and a commercial TelosB node to receive the LoRa, Wi-Fi, Bluetooth, and ZigBee signals generated by RF-Transformer, respectively. We measure the throughput and backscatter range of these commercial devices. Due to the limited indoor space, we report the outdoor experimental results of the LoRa node.

**Results.** Figure 16(b) and Figure 16(c) show the result. We observe that these commercial devices achieve comparable throughput and backscatter range with the software-defined radio platform USRP. The throughput of the LoRa, Wi-Fi 802.11b, Wi-Fi 802.11g/n/ac, Bluetooth, and ZigBee are 2.4 Kbps, 922.4 Kbps, 20.2 Mbps, 935.6 Kbps, and 201.5 Kbps, respectively. The backscatter range of the LoRa, Wi-Fi 802.11b, Wi-Fi 802.11g/n/ac, Bluetooth, and ZigBee are 516.5 m, 21.5 m, 14.8 m, 18.5 m, and 24.5 m, respectively. Compared with the USRP platform, the throughput drops by 1.12–1.35× and the backscatter range drops by 1.06–1.43 across all backscatter links. The reason of performance gap between the commercial device and the USRP mainly comes from the difference of hardware receiving sensitivity.

### 7.5 Proof-of-concept Application

RF-Transformer opens up a new form of ubiquitous communication for low-power IoT devices. In this section, we demonstrate a proof-of-concept application on foot traffic density monitoring indoors.

Foot traffic density monitoring is one of the key initiatives for contact tracing during pandemics (*e.g.*, COVID-19). We deploy tactile pressure sensors in corridors to detect human presence, as shown in Figure 17(a). RF-Transformer then modulates these sensor data by synthesizing LoRa signals on top of Wi-Fi signals – the most prevalent wireless signals indoors. The modulated LoRa signals are reflected back to a remote LoRa gateway for centralized data collection and processing. We compare the packet reception ratio (PRR) with a commercial LoRa node.

**Results.** As shown in Figure 17(b), we observe that RF-Transformer achieves comparable PRR with active LoRa nodes. The PRR of synthesized LoRa signals varies from 85.4% to 92.2%, 5.8%–10.6% lower than the standard LoRa node. The performance gap is expected since backscatter signal becomes very weak after two-way attenuation.

## 8 CONCLUSION

We have presented the design, implementation, and evaluation of RF-Transformer, a unified backscatter radio hardware abstraction. RF-Transformer provides a programmable interface to the microcontroller and allows IoT devices to synthesize different types of protocol-compliant backscatter signals. RF-Transformer also supports cross-technology backscatter that synthesizes LoRa packets on top of Wi-Fi signals. Field study shows that RF-Transformer achieves 23.8 Mbps, 247.1 Kbps, 986.5 Kbps, and 27.3 Kbps throughput when generating standard Wi-Fi, ZigBee, Bluetooth, and LoRa signals while consuming 7.6–74.2× less power than their active counterparts. We believe RF-Transformer is a significant step in a line of works that will scale out backscatter technologies to heterogeneous wireless networks. In the future, we plan to extend the design of RF-Transformer to frequency bands other than 2.4 GHz, as well as to explore the research space in cross-frequency backscatter.

## ACKNOWLEDGMENT

We thank our anonymous shepherd and reviewers for their insightful comments. This work is supported in part by National Science Fund of China under grant No. U21B2007, R&D Project of Key Core Technology and Generic Technology in Shanxi Province No. 2020XXX007, China Postdoctoral Science Foundation No. 2021M701888, and Key Program of China Postdoctoral Science Foundation No. 2022T150354. Dr. Longfei Shangguan is supported by his start-up funding from the University of Pittsburgh.

RF-Transformer:
A Unified Backscatter Radio Hardware Abstraction